\documentclass[prb,twocolumn,showpacs]{revtex4}
\usepackage{amsmath,amssymb,mathrsfs}
\usepackage{psfrag}
\usepackage{graphicx}
\usepackage{graphics}
\usepackage{epsfig}
\usepackage{bm}
\usepackage{color}
\usepackage{verbatim,color,ulem}

\newcommand{\beq}{\begin{equation}}
\newcommand{\eeq}{\end{equation}}
\newcommand{\beqa}{\begin{eqnarray}}
\newcommand{\eeqa}{\end{eqnarray}}

\begin{document}

\title{Finite-temperature quantum fluctuations 
in two-dimensional Fermi superfluids}
\author{G. Bighin$^{1,2}$ and L. Salasnich$^{1,3}$}
\affiliation{$^{1}$Dipartimento di Fisica e Astronomia ``Galileo Galilei'', 
Universit\`a di Padova, Via Marzolo 8, 35131 Padova, Italy
\\
$^{2}$Istituto Nazionale di Fisica Nucleare, Sezione di Padova, 
Via Marzolo 8, 35131 Padova, Italy
\\
$^{3}$Consorzio Nazionale Interuniversitario per le Scienze Fisiche 
della Materia (CNISM), Unit\`a di Padova, Via Marzolo 8, 35131 Padova, Italy
}

\date{\today}

% 3493 words

\begin{abstract}
In two-dimensional systems with a continuous symmetry
the Mermin-Wagner-Hohenberg theorem precludes spontaneous
symmetry breaking and condensation at finite temperature.
The Berezinskii-Kosterlitz-Thouless critical temperature marks 
the transition from a superfluid phase characterized by quasi-condensation
and algebraic long-range order to a normal phase, where vortex proliferation completely 
destroys superfluidity. As opposed to conventional 
off-diagonal long-range order typical of three-dimensional 
superfluid systems, 
algebraic long-range order is driven by quantum and thermal fluctuations 
strongly enhanced in reduced dimensionality. 
Motivated by this unique scenario and by the very recent experimental 
realization of trapped quasi-two-dimensional fermionic clouds, 
we include one-loop Gaussian fluctuations in the theoretical description 
of resonant Fermi superfluids in two dimensions demonstrating 
that first sound, second sound and also critical temperature 
are strongly renormalized, away from their mean-field values. 
In particular, we prove that in the intermediate and 
strong coupling regimes these quantities 
are radically different when Gaussian fluctuations are 
taken into account. Our one-loop theory shows good agreement 
with very recent experimental data on the 
Berezinskii-Kosterlitz-Thouless critical temperature 
[Phys. Rev. Lett. {\bf 115}, 010401 (2015)] and
on the first sound velocity,
giving novel predictions for the second sound 
as a function of interaction strength and temperature, 
open for experimental verification. 
\end{abstract}


\maketitle

\noindent {\it Introduction.}---Quantum fluctuations play a crucial role in low-dimensional systems, rendering the finite
temperature properties of a two-dimensional Fermi gas across the BCS-BEC crossover
substantially different from its three-dimensional counterpart.
In particular, in accordance with the Mermin-Wagner-Hohenberg theorem \cite{mermin,hohenberg,coleman} for $d \leq 2$ there can not
be a finite condensate density at finite temperature, as the fluctuations destroy the off-diagonal
long-range order; nonetheless two-dimensional systems can exhibit algebraic off-diagonal long-range
order, allowing for the existance of a quasi-condensante up to a certain critical temperature,
due to the Berezinskii-Kosterlitz-Thouless (BKT) mechanism \cite{berezinskii,kosterlitz}.

Along with the appearance of algebraic long-range order, as observed for the first time
in superfluid $^4\textrm{He}$, then in an ultracold Bose gas \cite{hadzibabic}, in an exciton-polariton gas \cite{nitsche}
and very recently in an ultracold Fermi gas \cite{murthy}, the other fundamental signature of the BKT mechanism
at work is the universal jump in the superfluid density \cite{nelson}, going discontinuously from a
finite value to zero at the critical temperature, as observed in thin $^4\textrm{He}$ films \cite{bishop}.
This scenario suggests that in a two-dimensional system the role of quantum fluctuations
should be crucial in describing several aspects of the system \cite{bertaina}, as opposed
to the 3D case for which one could expect from a mean-field theory at least
qualitative agreement.
Recently the strongly-interacting Fermi gas has been the object of numerous Montecarlo \cite{anderson,shi} and experimental \cite{orel,fenech,boettchereos} investigations, and in fact it has been observed that
Gaussian fluctuations strongly modify both the
chemical potential and the pairing parameter, particularly in the intermediate and strong
coupling regions \cite{he}; it has also been shown that the correct composite-boson limit is recovered by introducing Gaussian fluctuations \cite{salasnich3}.

The determination of a full one-loop Gaussian-level equation of state needs, however, a proper
regularization scheme to remove divergences. In the present Letter we use convergence factors
in the pair-fluctuation propagator \cite{diener,he} to numerically calculate the $T=0$ state equation for a
system of interacting fermions across the BCS-BEC crossover.
The aim of the present Letter is the investigation of 
beyond mean-field effects at finite temperature:
we calculate the first and second sound
velocities, as a function of the temperature and of the binding energy, and then calculate the
BKT critical temperature from the Kosterlitz-Nelson condition \cite{nelson}. In particular the predictions regarding the second sound velocity provide a benchmark for future experimental investigations: we expect it to be open to experimental verification quite soon, given the rapid advancements in the realization and manipulation of ultracold quasi-2D Fermi gases \cite{makhalov}. On the other hand the theoretical predictions regarding the BKT critical temperature and the first sound velocity are compared with recently obtained experimental results \cite{luick,murthy}, showing good agreement in the intermediate and BEC regimes.

{\it Theoretical framework.}---The partition function of a system of ultracold, dilute, interacting spin $1/2$ fermions in 2D,
contained in a two-dimensional volume $L^2$, at temperature $T$, with chemical potential $\mu$
can be described within the path-integral formalism \cite{nagaosa,stoof} as:
\beq
\mathcal{Z} = \int \mathcal{D} \psi_\sigma \mathcal{D} \bar{\psi}_\sigma e^{- \frac{1}{\hbar} \int_0^{\hbar \beta} \mathrm{d} \tau \int_{L^2} \mathrm{d}^2 r \mathscr{L}}
\label{eq:z}
\eeq
with the following (Euclidean) Lagrangian density
\beq 
\mathscr{L} = \bar{\psi}_{\sigma} \left[ \hbar \partial_{\tau} 
- \frac{\hbar^2}{2m}\nabla^2 - \mu \right] \psi_{\sigma} 
+ g \, \bar{\psi}_{\uparrow} \, \bar{\psi}_{\downarrow} 
\, \psi_{\downarrow} \, \psi_{\uparrow} \; , 
\label{lagrangian-initial}
\eeq
where $\psi_\sigma (\mathbf{r},\tau)$ and $\bar{\psi}_\sigma (\mathbf{r},\tau)$ are complex Grassmann fields, $\sigma=\uparrow,\downarrow$ is the spin index, $m$ is the mass of a fermion, having defined $\beta=(k_B T)^{-1}$, $k_B$ being the Boltzmann constant. The strength of the attractive s-wave potential is $g<0$, which can be implicitely related to the bound state energy \cite{randeria2,marini}:
\beq 
- \frac{1}{g} = \frac{1}{2L^2} \sum_{\bf k} \frac{1}{\epsilon_k + 
\frac{1}{2} \epsilon_B} \; . 
\label{g-eb}
\eeq
with $\epsilon_k = \hbar^2 k^2/(2m)$. In 2D, as opposed to the 3D case, a bound state exists even for arbitrarily weak interactions,
making $\epsilon_B$ a good variable to describe the whole BCS-BEC crossover. The quartic interaction can be decoupled by using a Hubbard-Stratonovich transformation in the Cooper channel, introducing in the process the new auxiliary pairing fields $\Delta (\mathbf{r},\tau)$, $\bar{\Delta} (\mathbf{r},\tau)$ \cite{nagaosa,stoof}. They correspond to a Cooper pair, being conjugate to two electron creation/annihilation operators \cite{altland}. Moreover the newly introduced pairing fields can be split into a uniform, constant saddle-point value $\Delta_0$ and the fluctuations around this value as follows:\beq 
\Delta({\bf r},\tau) = \Delta_0 +\eta({\bf r},\tau)  \; .
\label{polar}
\eeq

Neglecting the fluctuation fields $\eta$, $\bar{\eta}$ gives us a simple mean-field (MF) theory, which is 
generally unreliable, due to the fundamental role of fluctuations in two dimensions, but still constitutes the starting point for more refined approaches.
In the $T=0$ limit the functional integral and the $\mathbf{k}$-integrations defining the partition function at mean-field level are elementary, so that one finally gets the mean-field contribution to the equation of state \cite{gusynineos}, as derived in Appendix \ref{app:mf}:
\beq 
\Omega_{mf} (\mu) = - {m L^2\over 2\pi \hbar^2} 
(\mu + {1\over 2} \epsilon_B )^2  \; . 
\label{omega-mf}
\eeq
and the single particle excitations of the mean-field theory are:
\beq
E_k = \sqrt{(\epsilon_k - \mu)^2 + \Delta_0^2} \;.
\label{eq:ek}
\eeq
On the other hand one can extend the theory including the fluctuations fields at a Gaussian level \cite{diener,tempere}. The resulting equation of state reads:
\beq
\Omega_g (\mu)= {1\over 2\beta} \sum_{Q} \ln{\mbox{det}({\bf M}(Q))} \; ,
\label{eq:omegag}
\eeq
the full analytical expression of the inverse pair fluctuation propagator $\mathbf{M}(Q)$ is reported, along with a more detailed derivation, in the Appendix \ref{app:fl}. The collective excitations at $T=0$ are gapless due to Goldstone theorem \cite{goldstone}:
\beq 
\hbar \omega_q = 
\sqrt{\epsilon_q \left( \lambda \epsilon_q + 2 m c_s^2 \right)} 
\label{eq:col}
\eeq
At finite temperature a gap will appear, however it is extremely small below $T_{BKT}$
as shown in Ref. \cite{chien}.
From Eq. (\ref{omega-mf}) and Eq. (\ref{eq:omegag}) we get the full one-loop equation of state, by also using Eq. (\ref{g-eb}) we get $\mu$ as a function of the crossover. %The pairing gap $\Delta_0$ is found by inserting $\mu$ into Eq. (\ref{gapeq}).

In conclusion of the present section we note the grand-potential in Eq. (\ref{eq:omegag}) cannot be evaluated as is, being
affected by divergencies related to the modeling of the interaction using a contact pseudo-potential
rather than a realistic one. Many different regularization approaches can be used, like the dimensional regularization in 2D in the BEC limit \cite{salasnich3}, the counterterms regularization \cite{salasnich4} or regularization with convergence factors \cite{he,diener}. The first two are more suited to obtain analytical results, particularly in the BEC limit, while the last method has been shown to be suited to obtain numerical results across the whole crossover \cite{he}. Wanting to investigate numerically the whole crossover, our grand potential is regularized by introducing convergence factors \cite{diener,he}:
\beq
\Omega_g (\mu) = {1\over 2\beta} \sum_{Q} \ln \left[ \frac{\mathbf{M}_{11} (Q)}{\mathbf{M}_{22} (Q)} \mbox{det}({\bf M}(Q)) \right] e^{\mathrm{i} \Omega_n 0^+} \;.
\label{eq:omegagreg}
\eeq

\begin{figure}
\includegraphics[width=1.\linewidth,clip]{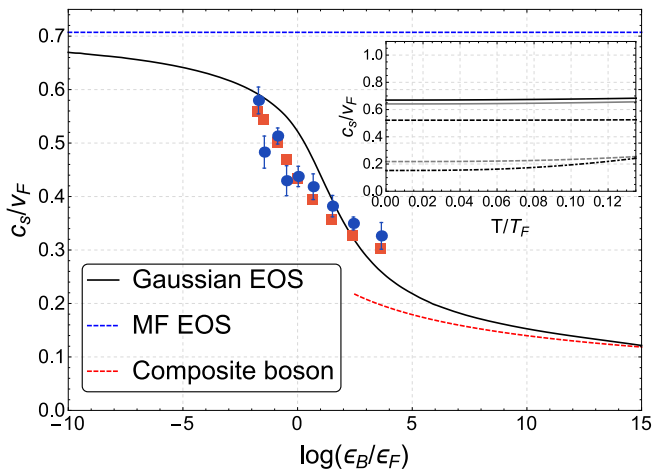}
\caption{The first sound velocity at $T=0$, calculated using $\mu$ and $\Delta_0$ from the Gaussian-level equation of state (black solid line), and using their mean-field counterparts (blue dashed line), which give a constant value $c_s/v_F = 1/\sqrt{2}$. In the strong coupling regime a full Gaussian-level equation of state is needed to correctly describe thermodynamic quantities, there our prediction correctly tends to the composite boson limit (red dotted line). Blue circles and red squares are experimental observations of the sound velocity, as reported in \cite{luick}, respectively obtained by directly measuring the speed of a density wave and through the equation of state. Inset: temperature dependence for $\log(\epsilon_B/\epsilon_F) = -10,-5,0,5,10$ (from top to bottom).} 
\label{fig1}
\end{figure}

{\it First and second sound.}---The first sound velocity $c_s$ is calculated from the regularized grand potential in Eq. (\ref{eq:omegagreg}), by using the zero-temperature thermodynamic relation \cite{lipparini}:
\beq
c_s = \sqrt{\frac{n}{m} \frac{\partial \mu}{\partial n}} = \sqrt{- \frac{n}{m} \left( \frac{1}{L^2} \frac{\partial^2 \Omega (\mu)}{\partial \mu^2} \right)^{-1}} \;.
\label{eq:cs}
\eeq
Using the mean-field equation of state to calculate the chemical potential, one would find $c_s(\mu_{mf})=v_F/\sqrt{2}$ across the whole BCS-BEC crossover, $v_F$ being the Fermi velocity \cite{marini,marini2}. Our equation of state with Gaussian fluctuations yields, as expected, a critically different $c_s$: it slowly tends to the aforementioned value in the BCS limit, showing, on the other hand, a remarkable difference in the intermediate and BEC regimes, tending to the composite boson limit derived in Ref. \cite{salasnich3}. We plot this result in Fig. \ref{fig1}, noting that it exhibits good agreement with the experimental data in Ref. \cite{luick}. By adapting the thermodynamic approach of Ref. \cite{salasnich1} we verified that the $T$-dependence of $c_s$ in the superfluid phase is very weak, see the inset of Fig. \ref{fig1}.

\begin{figure}
\includegraphics[width=1.\linewidth,clip, trim=0 0 0 0]{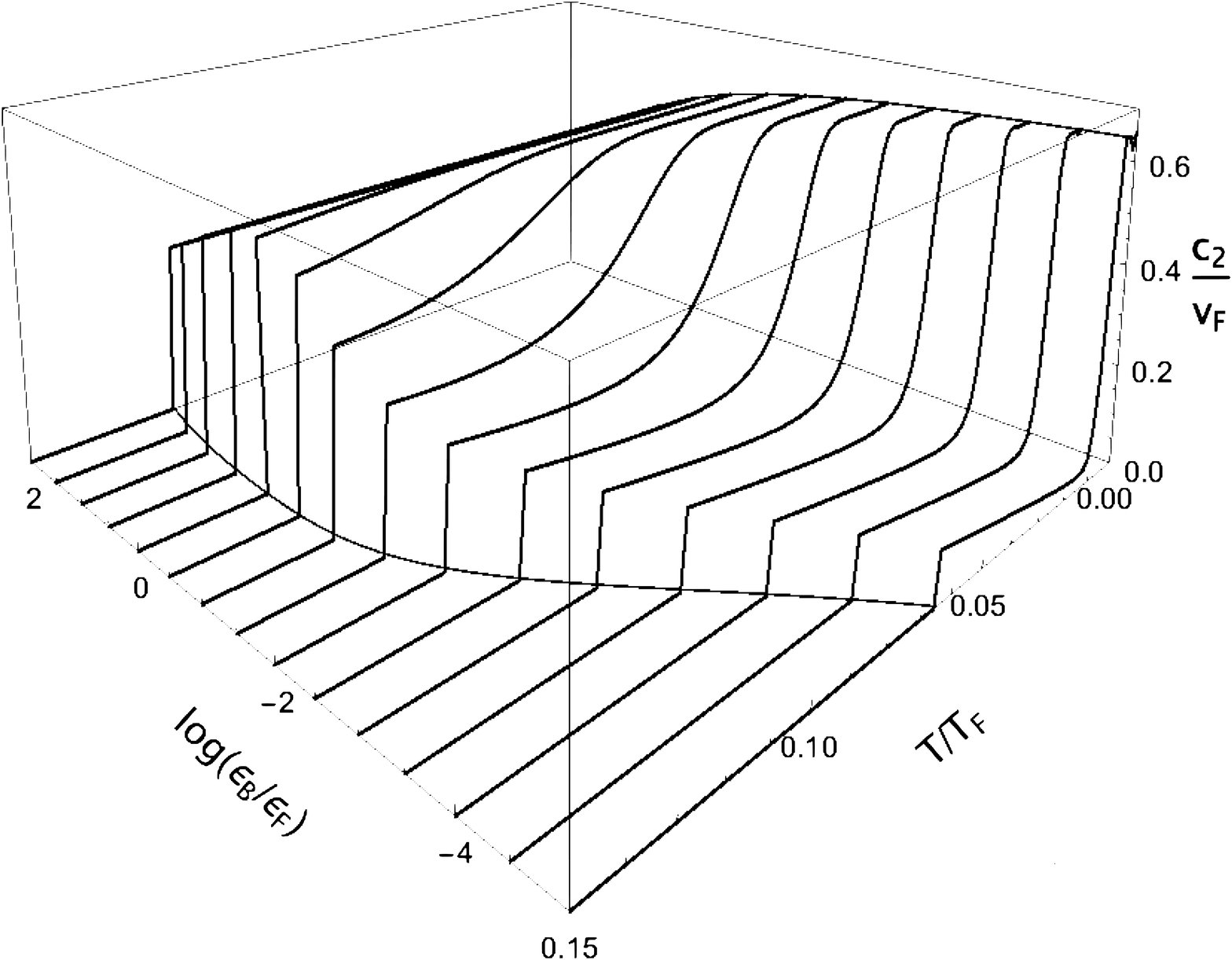}
\caption{The second sound velocity, as a function of the temperature $T/T_F$, for varying values of $\epsilon_B/\epsilon_F$. The characteristic structure with a minimum followed by a linear increase evolves into a constant second sound velocity approaching the BEC regime.} 
\label{fig2}
\end{figure}

Beside the first sound, propagating through density waves, a superfluid can also sustain the second sound, a purely quantum-mechanical phenomenon propagating through a temperature wave \cite{landau3}. In order to calculate the second sound velocity we follow the treatment in \cite{salasnich1} starting from the free energy of
the system, substantially treating it as a gas of independent single particle and collective excitations, neglecting hybridization through Landau damping; this approach will be justified shortly when discussing the BKT critical temperature. We find the fermion single particle contribution to the free energy\footnote{This approximate expression for the free energy is used only for calculating the second sound in the present Letter. We always use the full equation of state in Eq. (\ref{eq:omegagreg}) elsewhere.}:
\beq 
F_{sp} = - {2\over \beta} \sum_{\bf k} 
\ln{\left[ 1 + e^{-\beta E_k} 
\right]} \;
\label{eq:fsp}
\eeq
and the bosonic one, from collective excitations:
\beq 
F_{col} = {1\over \beta} \sum_{\bf q} 
\ln{\left[ 1 - e^{-\beta \omega_q} \right]} \;.
\label{eq:fcol}
\eeq
The total free energy is then $F=F_0+F_{col}+F_{sp}$ where the zero-temperature
energy $F_0$ is a $T$-independent constant, unimportant as far as the present Letter is
concerned. The entropy is readily calculated as $S=-(\partial F/\partial T)_{N,L^2}$ and introducing
the entropy per particle $\bar{S}=S/N$ the second sound velocity is \cite{landau,landau2,khalatnikov}:
\beq 
c_2 = \sqrt{{1\over m} {{\bar S}^2 \over 
\left({\partial {\bar S}\over \partial T}\right)_{N,L^2}} 
{n_s\over n_n} } \; .   
\label{c2}
\eeq
where $n_s$ and $n_n$ are the superfluid density and the normal fluid density, respectively.
In contrast with the 3D case \cite{salasnich1} here the second sound has a discontinuity at the
critical temperature, as a consequence of the universal jump in the superfluid density, as also noted in \cite{ozawa,liu}; the critical temperature will be calculated in the next section. Our results are reported in Fig. \ref{fig2}, we note that the second sound velocity shows a characteristic minimum in the BCS and intermediate regimes, as also noted in the 3D unitary case \cite{salasnich1}, evolving into an approximately constant second sound velocity approaching the BEC regime.

{\it Critical temperature: the Berezinskii-Kosterlitz-Thouless transition.}---As mentioned in the Introduction the low-temperature physics of a 2D attractive Fermi gas is essentially different from that of a 3D gas: the
Mermin-Wagner-Hohenberg theorem \cite{mermin,hohenberg,coleman} prohibits the
symmetry breaking at finite temperatures,
so that one can find off-diagonal long-range order and a finite condensate
density only at $T=T_c=0$. However quasi-condensation, i.e. the the algebraic decay
of the phase correlator $\langle \exp(\mathrm{i} \theta (\mathbf{r})) \exp(\mathrm{i} \theta (0)) \rangle \sim \left| \mathbf{r} \right|^{-\eta}$
where $\eta$ is a $T$-dependent exponent and $\theta$ is the phase of the order parameter,
is observed up to a finite temperature
$T_{BKT}$, known as the Berezinskii-Kosterlitz-Thouless (BKT) critical temperature
\cite{berezinskii,kosterlitz}.
The other fundamental signature of the BKT mechanism is the universal jump in superfluid density
at the critical temperature, i.e. $n_s (T_{BKT}^-) \neq (T_{BKT}^+)=0$.
The transition temperature is determined through the Kosterlitz-Nelson \cite{nelson} condition
\beq
k_B T_{BKT} = \frac{\hbar^2 \pi}{8 m} n_s (T_{BKT})
\label{eq:kncondition}
\eeq
which allows one to calculate $T_{BKT}$, known the superfluid density. Within the present framework we write the superfluid density as $n_s = n - n_{n,f} - n_{n,b}$
where $n$ is the density of the system and $n_{n,f}$ and $n_{n,b}$ are normal density contributions arising, respectively,
from the single particle excitations and from the bosonic collective excitations. Using Landau's quasiparticle excitations formula \cite{fetter} for fermionic:
\beq
n_{n,f} = \beta \int \frac{\mathrm{d}^2 k}{(2 \pi)^2} k^2 \frac{e^{\beta E_k}}{(e^{\beta E_k} + 1)^2}
\label{eq:nnf}
\eeq
and for bosonic excitations:
\beq
n_{n,b} = \frac{\beta}{2} \int \frac{\mathrm{d}^2 q}{(2 \pi)^2} q^2 \frac{e^{\beta \omega_q}}{(e^{\beta \omega_q} - 1)^2} \;.
\label{eq:nnb}
\eeq
The single particle excitation spectrum is $E_k = \sqrt{(k^2/(2m) - \mu)^2 + \Delta_0^2}$, as derived in Eq. (\ref{eq:ek}), the collective excitations spectrum in Eq. (\ref{eq:col}) can replaced by a phonon-like linear mode $\omega_q \sim c_s q$ within a very good approximation as far as the critical temperature is concerned, as we verified.

As noted in Ref. \cite{griffin} Eq. (\ref{eq:nnf}) and Eq. (\ref{eq:nnb}) hold as long as there is no Landau damping hybridizing the collective modes with fermionic single-particle excitations, otherwise the bosonic normal density would need to be modified. In our case one can easily verify that for $\epsilon_B \gtrsim 1$ the condition $\epsilon_B \gg k_B T$ holds in the whole temperature region of interest, strongly suppressing the pair breakup and the Landau damping \cite{griffin}. On the other hand, for $\epsilon_B \lesssim 1$ we verify that the critical temperature is determined by the fermionic contribution to the normal density, as one would expect, making eventual corrections to $n_{n,b}$ neglectable. We then conclude that Eq. (\ref{eq:nnf}) and Eq. (\ref{eq:nnb}) correctly describe the normal density for the entire superfluid phase.

\begin{figure}
\includegraphics[width=1.\linewidth,clip]{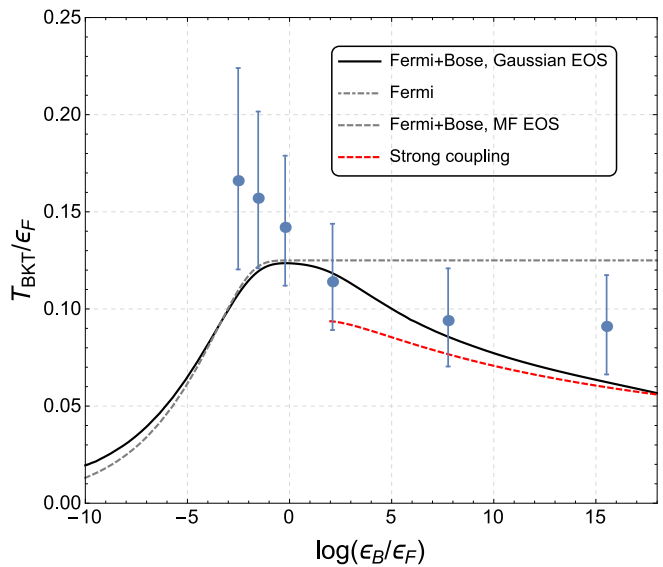}
\caption{Our theoretical prediction for $T_{BKT}$ (black solid line) as compared to recent experimental
observation reported in \cite{murthy}, temperature estimated through algebraic decay, the error bars account for statistical and systematic errors. Our
prediction uses a Gaussian equation of state, including the contribution from single particle modes
and collective excitations. A theory with fermionic only excitations (gray dot-dashed line) fails to
provide an agreement with experimental data in the BEC regime, whereas a theory using a
mean-field equation of state would underestimate $T_{BKT}$ in the weak coupling regime (gray
dashed line). The strong coupling extrapolation (red dotted line) in Eq. (\ref{eq:extra}) gives good
results in the strong and intermediate coupling regimes.}
\label{fig3}
\end{figure}

By numerically solving Eq. (\ref{eq:kncondition}) we find the transition temperature $T_{BKT}$ at different points of the BCS-BEC crossover. Our results, shown in Fig. \ref{fig3}, are compared with very recently obtained experimental data in Ref. \cite{murthy}, showing a excellent agreement with experimental data at least for $\epsilon_B / \epsilon_F \gtrsim 1$.

We stress that with respect to other derivations of $T_{BKT}$ in the 2D BCS-BEC crossover \cite{tempere,salasnich2,gusynintbkt,bauer} the present theoretical prediction of $T_{BKT}$ includes the contribution from a Gaussian-level equation of state along with the contribution from the bosonic collective excitations. These contributions are critical in correctly fitting experimental data, as clear from in Fig. \ref{fig3}. We find that a theory of fermionic only excitations, like those developed in \cite{tempere,salasnich2} or in a slightly different context in \cite{zinner}, overestimates the critical temperature in the intermediate and strong-coupling regimes. Conversely, not using a Gaussian equation of state underestimates the critical temperature in the BCS regime, see Fig. \ref{fig3}.

Moving towards the BCS side of the crossover, however, the agreement is slightly worse, the experimental $T_{BKT}$ being bigger than $0.125 \epsilon_F$;  by inserting into Eq. (\ref{eq:kncondition}) the relation $n = m/(\hbar^2 \pi) \epsilon_F$ it is easily seen that the critical temperature is not allowed to exceed the value $T_{BKT}=0.125 \epsilon_F$. % under the very general assumption that the superfluid density is a decreasing function of the temperature with $n_s(T=0)=n$.
Thus we conclude that the slightly worse compatibility observed cannot be reproduced within the framework of the Kosterlitz-Nelson criterion, as defined by Eq. (\ref{eq:kncondition}), and should be attributed to different physics, like the mesoscopic effects mentioned in \cite{makhalov} in the same regime. Nonetheless we stress that our results are still within $1.2 \sigma$ from experimental data, when statistical and systematic errors are taken into account.

{\it Strong coupling limit}---In the strong coupling limit an attractive Fermi gas maps into a Bose gas, in particular by using the relation between the fermionic and bosonic scattering lengths $a_B = a_F / (2^{1/2} e^{1/4})$ \cite{salasnich3} and the relation $\epsilon_B = 4 \hbar^2/(e^{2 \gamma} m a^2_F)$ \cite{mora} one finds $\epsilon_B/\epsilon_F = \kappa/(n_B a^2_B)$ with $\kappa=\exp(-2 \gamma - 1/2)/\pi \approx 0.061$ and $n_B = n_F / 2$. %The fundamental observation here is that
The strong coupling regime for the Fermi gas corresponds to the extremely dilute limit for the Bose gas.
In this limit the fermionic contribution to the normal
density in Eq. (\ref{eq:nnf}) is neglectable because the energy gap $\Delta_0$ becomes extremely large. Moreover the integration for $n_{n,b}$ in Eq. (\ref{eq:nnb}) is analytic, we solve the Kosterlitz-Nelson condition and expand in powers of $c_s$, obtaining the following analytical estimate for the critical temperature in the strong-coupling regime:
\beq
k_B T_{BKT} \approx \frac{\mu_B^{\frac{2}{3}} \epsilon_F^{\frac{1}{3}}}{\sqrt[3]{6 \zeta(3)}} - \frac{2^\frac{1}{3} 4}{3} \frac{\mu_B^\frac{4}{3} \epsilon_F^{-\frac{1}{3}}}{(3 \zeta(3))^\frac{2}{3}}
\label{eq:extra}
\eeq
with $\zeta(3) \approx 1.202$ and $\mu_B = \epsilon_F/\ln{\left({1/(n_B a_B^2)}\right)}$ as in Ref. \cite{salasnich3}, we report this result in Fig. \ref{fig3}. An alternative estimate of $T_{BKT}$ can be derived mapping the attractive Fermi gas to a Bose gas, we report the derivation in Appendix \ref{app:alt}, noting however that his validity is limited to the strong coupling regime.

{\it Conclusions.}---There are several open problems for the physics of ultracold atoms 
which can be faced employing one-loop Gaussian fluctuations. 
Here we have shown that the Berezinsky-Kosterlitz-Thouless critical 
temperature of the superfluid-normal phase transition can be extracted 
from an description of the superfluid density, 
which takes into account Gaussian fluctuations in 
the finite-temperature equation of state. The agreement  
with very recent experimental data for both the critical temperature \cite{murthy}
and the sound velocity \cite{luick} is remarkably good and 
crucially depends on the inclusion of quantum and thermal Gaussian 
fluctuations. More generally, Gaussian contributions to the equation of state 
are relevant for Bose-Fermi mixtures \cite{nishida2006}, 
for unbalanced superfluid fermions \cite{klimin2012}, and also to investigate 
the condensate fraction in the BCS-BEC crossover \cite{griffin}. 
Finally, we stress that in addition to ultracold atomic gases  
there are several other superfluid quantum many-body systems 
where the methods of functional integration and Gaussian fluctuations 
play a relevant role to achieve a meaningful and reliable 
theoretical description. Among them we quote 
high-T$_c$ superconductors \cite{scalapino2012},  
neutron matter in the BCS-BEC crossover \cite{sala-nuclear}, 
quark-gluon plasma \cite{bhattacharya2014}, quark matter 
in stars \cite{anglani2014}, and, more generally, quantum fluids 
of light \cite{carusotto2013}. In particular, our results pave 
the way for a better understanding of the strong-coupling limit of other 
two-dimensional systems with BCS pairing, 
e.g. bilayers of fermionic polar molecules \cite{baranov,zinner} 
or exciton-polariton condensates \cite{byrnes}. 

\begin{acknowledgments}
The authors acknowledge Ministero Istruzione
Universita Ricerca (PRIN project 2010LLKJBX) for partial support.
The authors thank A. Perali, C. S\'a de Melo, G. Strinati, and 
V. Vukoje for fruitful discussions. 

\end{acknowledgments}

\appendix
\section{Mean-field treatment}
\label{app:mf}
\noindent Starting from the same Lagrangian as in the main text:
\beq
\mathscr{L} = \bar{\psi}_{\sigma} \left[ \hbar \partial_{\tau} 
- \frac{\hbar^2}{2m}\nabla^2 - \mu \right] \psi_{\sigma} 
+ g \, \bar{\psi}_{\uparrow} \, \bar{\psi}_{\downarrow} 
\, \psi_{\downarrow} \, \psi_{\uparrow} \; 
\eeq
with the same notation, after the Hubbard-Stratonovich transformation one obtains the new (Euclidean) Lagrangian density:
\beq 
\mathscr{L}_e =
\bar{\psi}_{\sigma} \left[  \hbar \partial_{\tau} 
- {\hbar^2\over 2m}\nabla^2 - \mu \right] \psi_{\sigma} 
+ \bar{\Delta} \, \psi_{\downarrow} \, \psi_{\uparrow} 
+ \Delta \bar{\psi}_{\uparrow} \, \bar{\psi}_{\downarrow} 
- {|\Delta|^2\over g}
\label{ltilde}
\eeq 
and the functional integration needs to be extended over $\Delta$, $\bar{\Delta}$. As mentioned in the main text the newly introduced pairing field can be split into a uniform, constant saddle-point value $\Delta_0$ and the fluctuations around this value as follows:
\beq 
\Delta({\bf r},\tau) = \Delta_0 +\eta({\bf r},\tau)  \; , 
\label{app:polar}
\eeq
The mean-field approximation consists in neglecting the the fluctuation fields $\eta$, $\bar{\eta}$; in this case the functional integral defining the partition function can be carried out exactly, yielding:

\beq 
{\cal Z}_{mf} =  \exp{\left\{ - {S_{mf}\over \hbar} \right\}}
= \exp{\left\{ - \beta \, \Omega_{mf} \right\}} \; , 
\eeq
where 
\beqa 
{S_{mf}\over \hbar} = - Tr[\ln{(G_0^{-1})}] - 
\beta {L^2} {\Delta_0^2\over g} = \nonumber \\
= - \sum_{{\bf k}} \left[ 2
\ln{\left( 2 \cosh{(\beta E_{sp}(k)/2)} \right)} 
- \beta \xi_\mathbf{k} \right] - \beta L^2 {\Delta_0^2\over g} \; , 
\label{omega-sp} 
\eeqa
with $\xi_\mathbf{k}=\epsilon_k -\mu$, the trace being taken in reciprocal space and in the Nambu-Gor'kov space, with
\beq 
G_0^{-1} = \left(
\begin{array}{cc}
\hbar \partial_{\tau} -{\hbar^2\over 2m}\nabla^2 -\mu & \Delta_0 \\ 
\Delta_0 & \hbar \partial_{\tau} +{\hbar^2\over 2m}\nabla^2 +\mu
\end{array}
\right)
\label{G0}
\eeq
and the single-particle excitation spectrum is found solving for the poles of the Nambu-Gor'kov Green's function $G_0$ in momentum space \cite{stoof}:
\beq
E_k = \sqrt{(\epsilon_k - \mu)^2 + \Delta_0^2} \;.
\label{eq:app:ek}
\eeq
In the $T=0$ limit the $\mathbf{k}$-integration in Eq. (\ref{omega-sp}) can be carried out analytically in the 2D case, one gets:
\beqa 
\Omega_{mf} (\mu, \Delta_0) = - {m L^2\over 4\pi \hbar^2} \Big[ 
\mu^2 + \mu \sqrt{\mu^2+\Delta_0^2} + {1\over 2} \Delta_0^2 + \nonumber \\
- \Delta_0^2 \ln{\Big({-\mu + \sqrt{\mu^2 + \Delta_0^2} 
\over \epsilon_B}\Big)} \Big]  \; . 
\label{omega-mf-full}
\eeqa
Imposing the saddle-point condition for $\Delta_0$,
i.e. $(\partial \Omega_{mf} / \partial \Delta_0)_{\mu,V} = 0$, one obtains the gap equation:
\beq
\Delta_0 = \sqrt{2\epsilon_b ( \mu + {1\over 2}\epsilon_B ) } \; , 
\label{gapeq}
\eeq 
plugging this result back into the MF grand potential we get the MF equation of state:
\beq 
\Omega_{mf} (\mu) = - {m L^2\over 2\pi \hbar^2} 
(\mu + {1\over 2} \epsilon_B )^2  \; . 
\label{app:omega-mf}
\eeq

\section{Gaussian fluctuations}
\label{app:fl}

Restoring the fluctuation fields $\eta$, $\bar{\eta}$ at a Gaussian level, the partition
function reads \cite{diener}:
\beq 
\mathcal{Z} = \mathcal{Z}_{mf} \ \int 
{\cal D}\eta {\cal D}\bar{\eta} \ 
\exp{\left\{ - {S_g(\eta,\bar{\eta}) \over \hbar} \right\}} \; , 
\label{sigo}
\eeq
where 
\beq 
S_{g}(\eta,\bar{\eta}) = {1\over 2} \sum_{Q} 
({\bar\eta}(Q),\eta(-Q)) \ {\bf M}(Q) \left(
\begin{array}{c}
\eta(Q) \\ 
{\bar\eta}(-Q) 
\end{array}
\right) \;
\label{eq:sg} 
\eeq
having introduced the Fourier-transformed version of the fluctuation fields, with $Q=(\mathrm{i} \Omega_n, \mathbf{q})$, $\Omega_n = 2 \pi n / \beta$ being the Bose Matsubara frequencies.
The $2 \times 2$ matrix in Eq. (\ref{eq:sg}) is the inverse propagator for the pair fluctuations, its matrix elements are defined by \cite{diener,tempere}:
\beqa
\mathbb{M}_{11} = \frac{1}{g} + \sum_\mathbf{k} \left( \frac{u^2 u'^2}{\mathrm{i} \omega_n - E - E'} - \frac{v^2 v'^2}{\mathrm{i} \omega_n + E + E'} \right) \\
\mathbb{M}_{12} = \sum_\mathbf{k} u v u' v' \left( \frac{1}{\mathrm{i} \omega_n + E + E'} - \frac{1}{\mathrm{i} \omega_n - E - E'} \right)
\eeqa
where $u=u_{\mathbf{k}} = \sqrt{\frac{1}{2} ( 1 + \frac{\epsilon_\mathbf{k} - \mu}{E_{sp} (\mathbf{k})})}$, $v=v_\mathbf{k}=\sqrt{1-u^2_{\mathbf{k}}}$, $u'=u_{\mathbf{k}+\mathbf{q}}$, $v'=v_{\mathbf{k}+\mathbf{q}}$, $E=E_{sp} (\mathbf{k})$, $E'=E_{sp} (\mathbf{k} + \mathbf{q})$. The remaining matrix elements are defined by the relations: $ \mathbb{M}_{22} (q)  = \mathbb{M}_{11} (-q) $, $ \mathbb{M}_{21} (q)  = \mathbb{M}_{12} (q) $. The quantity $1/g$ appearing in the definition of $\mathbb{M}_{11} (Q)$ is removed by using the scattering theory result:
\beq 
- \frac{1}{g} = \frac{1}{2L^2} \sum_{\bf k} \frac{1}{\epsilon_k + 
\frac{1}{2} \epsilon_B} \; . 
\label{app:g-eb}
\eeq
By integrating out the $\eta(\mathbf{r},\tau)$, $\bar{\eta}(\mathbf{r},\tau)$ fields in Eq. (\ref{sigo}) we get the Gaussian contribution to the grand potential:
\beq
\Omega_g (\mu, \Delta_0) = \frac{1}{2 \beta} \sum_q \ln ( \det \mathbb{M} (q))
\eeq
and the equation of state $\Omega_g (\mu)$ is found, like in the MF case, by inserting the gap equation from Eq. (\ref{gapeq}).
By imposing the condition $\det (\mathbf{M}) = 0$ one can find the collective excitation spectrum, which will have, in the low-momentum limit, the following expression
\beq 
\hbar \omega_q = 
\sqrt{\epsilon_q \left( \lambda \epsilon_q + 2 m c_s^2 \right)} 
\label{eq:app:col}
\eeq
$\lambda$ and $c_s$ being a function of the crossover. As already noted in the main text the collective excitation spectrum is gapless as a consequence of Goldstone theorem.

\section{Alternative determination of the critical temperature in the Bose limit}
\label{app:alt}

An alternative estimate of the BKT transition temperature can be given by extending to the intermediate coupling regime the strong coupling result valid for $\epsilon_B \gg \epsilon_F$. As noted in the main text in this limit an attractive Fermi gas maps to a Bose gas, each boson has mass $m_B = 2 m_F$ and the density is $n_B = n_F/2$. The Fermi energy of the original Fermi gas is then:
\beq
\epsilon_F = \frac{\hbar^2 \pi}{m_F} n_F = 4 \frac{\hbar^2 \pi}{m_B} n_B
\label{eq:efnb}
\eeq
From now to the end of the present section $F$ and $B$ subscripts will be used to distinguish between, respectively, fermionic and bosonic masses, densities and scattering lengths. The binding energy $\epsilon_B$ and the fermionic scattering length $a_F$ are related by the equation found by Mora and Castin \cite{mora}:
\beq
\epsilon_B = \frac{4 \hbar^2}{e^{2 \gamma} m a^2_F}
\label{eq:moracastin}
\eeq
It can combined with the relation found between the bosonic and fermionic scattering lengths \cite{salasnich3} $a_B = a_F / (2^{1/2} e^{1/4})$ to obtain:
\beq
\frac{\epsilon_B}{\epsilon_F} = \frac{\kappa}{n_B a^2_B}
\label{eq:ebgp}
\eeq
where $\kappa=\exp(-2 \gamma - 1/2)/\pi \approx 0.061$, $\gamma$ being the Euler-Mascheroni constant. The Berezinskii-Kosterlitz-Thouless critical temperature for a dilute 2D Bose gas \cite{fisher} has been estimated using Montecarlo techniques \cite{posazhennikova,prokofev}:
\beq
\frac{T_{BKT}}{n_B} = \frac{2 \pi}{m_B \log(\frac{\xi}{m_B U_\text{eff}})}
\label{eq:prokofev}
\eeq
where
\beq
U_\text{eff}=\frac{4 \pi}{m_B \log (1/na_B^2)}
\eeq
and Montecarlo simulations yield $\xi \sim 380$. Moreover Eq. (\ref{eq:prokofev}) can be rewritten by using Eq. (\ref{eq:efnb}) and Eq. (\ref{eq:ebgp}) as:
\beq
\frac{T_{BKT}}{\epsilon_F} = \frac{1}{2} \frac{1}{\log ( \frac{\xi}{4 \pi} \log ( \kappa^{-1} \frac{\epsilon_B}{\epsilon_F}))}
\label{eq:tkt}
\eeq
This result can be compared with the experimental data reported in \cite{murthy}; we observe that the theoretical prediction in Eq. (\ref{eq:tkt}) correctly fits experimental data within the reported statistical errors. Conversely we can leave $\xi$ as a free parameter and estimate it through Eq. (\ref{eq:tkt}) and the experimental data using a simple least squares method, the result is $\xi = 554 \pm 179$ which is compatible with the Montecarlo estimate in \cite{prokofev}. However we must stress that this alternative result in Eq. (\ref{eq:tkt}) is essentially a composite-boson extrapolation from the strong coupling regime which is not reliable as we approach the BCS side of the resonance, as demonstrated by the divergence in $T_{BKT}$ when $\epsilon_B / \epsilon_F = \exp (4 \pi/\xi) \kappa \sim 0.063 $.

\section{Condensate fraction}
\label{app:n0}

\begin{figure}
\includegraphics[width=.78\linewidth,clip]{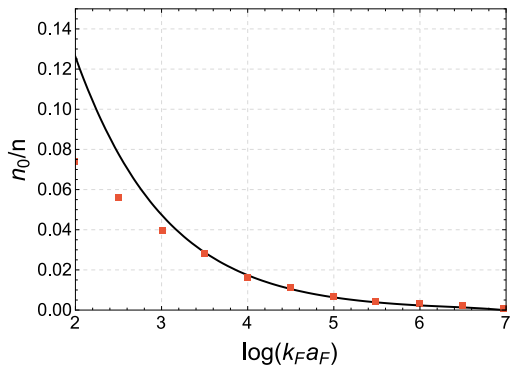}
\caption{The condensate fraction in the BCS and intermediate regimes, as calculated from Eq. (\ref{eq:cf}), compared with Montecarlo data from Ref. \cite{shi}.}
\label{smfig1}
\end{figure}

A fundamental quantity in the study of ultracold systems is the condensate fraction. In a 2D system one expects a finite condensate density $n_0$ only a $T=0$ due to the Mermin-Wagner-Hohenberg \cite{mermin,hohenberg} theorem. In the case of a 2D attractive Fermi gas the condensate density is given by \cite{fukushima}
\beq
n_0 = \lim_{\beta \to \infty} \frac{1}{L^2 \beta^2} \sum_{\mathbf{p},m,n} G_{21} (\mathbf{p}, \mathrm{i} \omega_m) G_{12} (\mathbf{p}, \mathrm{i} \omega_n)
\label{eq:n0}
\eeq
where $G$ is the single-particle Green's function, $\omega_n$, $\omega_m$ are Fermi Matsubara frequencies, and the condensate fraction is simply $n_0/n$. In the BCS limit all the integrals can be carried out analytically as in Ref. \cite{salasnich7}, yielding
\beq
\frac{n_0}{n} = \frac{1}{4} \frac{\frac{\pi}{2} + \arctan \left( \frac{\mu}{\Delta_0} \right)}{\frac{\mu}{\Delta_0} + \sqrt{1+ \left( \frac{\mu}{\Delta_0} \right)^2}} \; .
\label{eq:cf}
\eeq
We use the Gaussian equation of state for $\mu$ and $\Delta_0$ to compare this theoretical prediction for the condensate fraction with the recent Montecarlo results in Ref. \cite{shi}, as reported in Fig. \ref{smfig1}, in the BCS and intermediate regimes. Clearly the rather good agreement breaks as the interaction gets stronger; an extension of this analysis to the BEC regime would require the evaluation of Eq. (\ref{eq:n0}) across the crossover, recalling that for self-consistency the single-particle Green's function is to be calculated at one-loop level \cite{fukushima}. This calculation is the subject of ongoing work by the authors.

\end{document}